# Left-eigenvectors are certificates of the Orbit Problem.

Steven de Oliveira[1], Virgile Prevosto[1], Peter Habermehl[2], and Saddek Bensalem[3]

1   CEA, List
2   IRIF, Université Paris Diderot - Paris 7
3   Université Grenoble Alpes

## Abstract

This paper investigates the connexion between the Kannan-Lipton Orbit Problem and the polynomial invariant generator algorithm PILA based on eigenvectors computation. Namely, we reduce the problem of generating linear and polynomial certificates of non-reachability for the Orbit Problem for linear transformations with coefficients in $\mathbb{Q}$ to the generalized eigenvector problem. Also, we prove the existence of such certificates for any transformation with integer coefficients, which is not the case with rational coefficients.

## 1   Introduction

Finding a suitable representation of the reachable set of configurations for a given transition system or transformation is a fundamental problem in computer science, notably in program analysis and verification. An exact representation of the reachable set can generally not be exactly computed. In this context, invariants often provide a good balance between precision, conciseness and ease of use. Model-checking [12] and deductive verification [8] often require the user to provide invariants in order to reach a given proof objective. In practice, for large programs, manually writing each invariant for each loop is extremely costly and becomes quickly infeasible. Users can rely on invariants synthesizers, that manage to infer an over-approximation of the reachable set of configurations. Abstract interpretation [3, 1] for example is based on the propagation of abstract values, such as *e.g.* intervals or octagons, that encompass the whole set of possible concrete inputs. Dynamic inference [6] tries to infer a candidate invariant satisfied by a large amount of runtime executions. The quality of the synthesis is here dependent of the choosen invariant pattern. Mathematical properties of specific kinds of transformations, such as the use of linear algebra properties [2, 4] or the search of algebraic dependencies [11] can elegantly facilitate the automated search for invariants. For all of these techniques, the following issues arise:

1. they work under very specific hypotheses;
2. generated invariants may not be precise enough to succeed in proving or *disproving* a given property.

As an example, [4] and [5] describe the PILA method for generating invariants of linear transformations based on the eigenspace problem. This method relies on the stability of left-eigenvectors of a linear transformation: a left-eigenvector $\varphi$ of a linear transformation $f$ verifies $\varphi \circ f = \lambda \varphi$ for some constant $\lambda$. Depending on the value of $\lambda$, $\varphi$ leads to inductive

| Hypotheses on matrix $A$ with eigenvalue $\lambda$ | **Hypothesis 1** $\|\lambda\| \neq 0 \wedge \|\lambda\| \neq 1$ | **Hypothesis 2** $A$ not diagonalizable $\|\lambda\| = 1$ | **Hypothesis 3** $A$ diagonalizable $\|\lambda\| = 1$ |
|---|---|---|---|
| PILAT [4, 5] | Inequality invariants $P(X) \leqslant 0, P(X) \geqslant 0$ | Equality invariants $P(X) = 0$ | Equality invariants $P(X) = 0$ |
| [7] on the existence of certificates | General existence of a semialgebraic certificate | General existence of a semialgebraic certificate | Necessary & sufficient conditions for the existence of a semialgebraic certificate |
| Contributions | <ul><li>Existence of $M$ computing the same image than $A$</li><li>Eigenvectors of $M$ are certificates</li></ul> | <ul><li>Existence of $M$ computing the same image than $A$</li><li>Generalized eigenvectors of $M$ are certificates</li></ul> | <ul><li>Eigenvectors can be used as certificates under the same conditions</li></ul> |

■ **Table 1** Comparaison between PILAT, the results of [7] and the contributions of this paper.

invariants. For instance, if $\lambda = 1$, then $\forall X, \varphi \circ f(X) = \varphi(X)$. Though the method is complete for a certain shape of invariants (polynomial equalities $P(X) = 0$ and inequalities $|P(X)| \leqslant k$ up to a given degree $n$), it is not stated nor clear what is achievable thanks to those invariants and what is not.

**The Kannan-Lipton Orbit Problem.**

A particular instance of the reachability problem is called the *Kannan-Lipton Orbit Problem* [9, 10], which can be stated as follows :

Given a square matrix $A \in \mathcal{M}_d(\mathbb{Q})$ of size $d$ and
two vectors $X, Y \in \mathbb{Q}^d$, determine if there exists $n$ such that $A^n X = Y$.

This problem is decidable in polynomial time. In the case an instance of the problem has no solution (in other words, $Y$ is not reachable from $X$), [7] studies the existence of non-reachability semialgebraic certificates for a given instance of the Orbit Problem where $Y$ is not reachable. Semialgebraic certificates are sets described by conjunction and disjuction of polynomial inequalities with integer coefficients that include the reachable set of states but not the objective $Y$. Those certificates allow to quickly prove the non-reachability of the given vector $Y$ and all vectors outside of the certificate. [7] concludes on the existence of such certificates under simple hypotheses on the eigenvalue decomposition of $A$. Those hypotheses are surprisingly similar to the hypotheses of PILA, where the shape of the generated invariants strongly depends on eigenvalues as well.

In this paper we investigate the connexions between the construction of certificates for the Orbit Problem and the invariants generated by PILA as summarized in Table 1. We show that for an instance of the Orbit Problem for the transformation $A$ of dimension $n$, the problem of generating a certificate can be reduced to the search of eigenvectors. Particularly,

- in the first hypothesis, there exists a linear transformation of dimension $O(n^2)$ (resp. $O(2^n)$) computing an equivalent image of $A$ s. t. its eigenvectors can be used as real certificates (resp. semialgebraic certificates) for the non reachability of the given instance;

- in the second hypothesis, there exists a linear transformation of dimension $O(n^2)$ (resp. $O(2^n)$) computing an equivalent image of $A$ such that its *generalized* eigenvectors can be used as real certificates (resp. semialgebraic certificates) for the non reachability for the given instance;
- in a more general case, a semialgebraic certificate for the Orbit Problem in $\mathbb{Z}$ always exists.

It is worth noting that, to our knowledge, there exists no proof about the decidability of the existence of linear certificates directly on the transformation $A$.

## 2 Setting

Let $\mathbb{K}$ be a field and $d \in \mathbb{N}$. Given two vectors $u, v$ of same dimension, we note $\langle u, v \rangle = u^t.v$, with . the usual dot product (i.e. the sum of the product of each component of $u$ and $v$). Every linear transformation $f : \mathbb{K}^d \to \mathbb{K}^d$ corresponds to a square matrix $A_f \in \mathcal{M}_d(\mathbb{K})$. For any vector $\varphi \in \mathbb{K}^d$, $\varphi^t : \mathbb{K}^d \to \mathbb{K}$ will denote a linear transformation. When the context is clear, we will refer to $A_f$ as $A$. The application $f^* : (\mathbb{K}^d \to \mathbb{K}) \to (\mathbb{K}^d \to \mathbb{K})$ is called the *dual* of $f$ ($f^*(\varphi) = \varphi^t \circ f$). It is also a linear transformation, and its associated matrix is $A_{f^*} = A_f^t$ the transpose of $A_f$. The application obtained by $n$ successive applications of a transformation $f : \mathbb{K}^d \to \mathbb{K}^d$ is denoted by $f^n$ and its matrix is $A_f^n$. Affine transformations can be considered as linear transformation by adding an extra constant variable $\mathbb{1}$. For example, the transformation $f(x) = x+1$ can be considered equivalent to the transformation $g(x, \mathbb{1}) = (x + \mathbb{1}, \mathbb{1})$. In this way, every affine transformation also admits a unique matrix representation.

▶ **Definition 1.** Let $f : \mathbb{K}^d \to \mathbb{K}^d$ be a linear transformation. Then, $\varphi \in \mathbb{K}^d$ is called a $\lambda$-right-eigenvector (resp. $\lambda$-left-eigenvector) and $\lambda$ its corresponding eigenvalue if $f(\varphi) = \lambda\varphi$ (resp. $f^*(\varphi) = \lambda\varphi$).

When a concept can be applied to either left or right-eigenvectors, we will simply refer to them as eigenvectors.

▶ **Definition 2.** A family of linked generalized $\lambda$-eigenvectors $\mathcal{F}_f = \{e_0, ..., e_k\}$ for the transformation $f$ are vectors verifying for all $i \leqslant k$ :
- $f(e_0) = \lambda e_0$
- $f(e_i) = \lambda e_i + e_{i-1}$

**The Orbit Problem.** This article focuses on $\mathbb{A} \subset \mathbb{C}$, the field of algebraic numbers. Elements of $\mathbb{A}$ are roots of polynomials with integer coefficients. Indeed, the linear transformations we consider are in $\mathbb{Q}^d \to \mathbb{Q}^d$, thus their eigenvalues (as roots of the characteristic polynomial) are in $\mathbb{A}$. Let $f : \mathbb{Q}^d \to \mathbb{Q}^d$ be a linear transformation. We refer to the Orbit Problem of $A_f$ with an initial state $X \in \mathbb{Q}^d$ and an objective state $Y \in \mathbb{Q}^d$ as $\mathcal{O}(A, X, Y)$. In other words, $\mathcal{O}(A, X, Y) = (\exists n \in \mathbb{N}.Y = A^n X)$. As we are studying non-reachability, every instance of the problem is assumed to be false unless stated otherwise.

▶ **Definition 3.** A *non-reachability certificate* or just *certificate* is a couple $(N, P) \in \mathbb{N} \times \mathcal{P}(\mathbb{Q}^d)$ of an instance $\mathcal{O}(A, X, Y)$ such that :

- $\forall n \in \mathbb{N}, n < N \Rightarrow A^n X \neq Y$
- $\forall n \in \mathbb{N}, n \geqslant N \Rightarrow A^n X \in P$
- $Y \notin P$

$N$ is called the *certificate index* and $P$ the *certificate set*.

When the certificate set is described by a combination of linear (resp. polynomial) relations between variables of $X$, the certificate is called linear (resp. polynomial). Irrational, semi-algebraic and rational certificates are linear or polynomial certificates whose coefficients are respectively irrationals, algebraic integers or rationals.

Semi-algebraic certificates, are always equivalent to rational certificates. Indeed, every coefficient $\varphi_i \in \mathbb{A}$ is nullified by a polynomial $Q$ with integer coefficients. It is then possible to replace $\varphi_i$ by a free variable that is constrained to be a root of $Q$. For example, $P = \{x | \sqrt{2}x \leqslant 2\} = \{x | \exists y. y^2 = 2 \wedge y \geqslant 0 \wedge yx \leqslant 2\}$.

**Remarks.** The certificate sets we generate are *future invariants* of the transformation, in the sense that $f^n(X)$ eventually reaches the set for some $n$ and always remains in it, whereas $Y$ is outside the invariant. Different choices of $X$ and $Y$ may delay the number of iterations needed to reach it. The certificate index solves this issue by expressing the number of iterations necessary for $f^n(X)$ to reach the certificate set. This information is crucial for the practical use of certificates, as a solver can use it to shorten its analysis.

The existence of such a couple implies the non reachability of $Y$ as $A^n X$ is either different from $Y$ or belongs to a set to which $Y$ does not. For example, if $Y$ does not belong to the reachable set of states $R = \{A^n X \mid n \geqslant 0\}$, the couple $(0, R)$ is a certificate. However, typically, $R$ can not be described in a *non-enumerative* way. We are interested in *simple* certificates, i.e. where proving that the objective $Y$ does not belong to the reachable set of states is straightforward. That means that membership in $P$ should be easy to solve. For example, let $R' = \{(v_1, ..., v_n) | v_1 + v_2 \geqslant 0\}$ and assume $R \subset R'$. Testing whether $Y$ is in $R'$ or not is easy as this set is described by a linear combination of variables of $V$. If $Y \notin R'$, then $R'$ is generally a *better* (simpler) certificate set than $R$. On the other hand, finding a good certificate index may be harder. Its search is studied in section 3.1.

## 3 Invariants by generalized eigenvectors

### 3.1 Certificate sets of the rational Orbit Problem

The decidability of the existence or the non-existence of semialgebraic certificates for the Orbit Problem for rational linear transformations is proven in [7]. It classifies four categories of rational linear transformations $f : \mathbb{Q}^d \to \mathbb{Q}^d$:

- $f$ admits null eigenvalues;
- $f$ has at least an eigenvalue of modulus strictly greater or less than 1;
- $f$ has all its eigenvalues of modulus 1, but it is not diagonalisable;
- $f$ has all its eigenvalue of modulus 1 and is diagonalisable.

In the second and third case, linear transformations always admit a non reachability certificate if the Orbit problem has no solution. The intuition behind this result is to consider the Jordan normal form $f_J$ of the transformation $f$. Let $V$ be a vector of variables and $V_J$ the vector of variables in the base of $J$. In this form, there exists a variable $v_J$ (representing a linear combination of variables of $V$) such that $f_J(V_J)_{|_{v_J}} = \lambda v_J$. Applied $k$ times, the new value of $v_J$ is $\lambda^k v_J$, which diverges towards infinity or converges towards 0 when $|\lambda| \neq 1$. Checking if a value $y$ is reachable or not can then be done by checking if there exists $k \in \mathbb{N}$ such that $\lambda^k v_J = y$. We are now left to compute those certificates.

**Case 1: there exist null eigenvalues**

This particular case leads to degenerate instances of the orbit problem. When a linear transformation admits a null eigenvalue, there exists a linear combination of variables that is always null. In other words, there exists a variable $v$ that can be expressed as a linear combination of the other variables. Therefore, this variable doesn't provide any useful information on the transformation other than an easily checkable constraint on $v$. If the linear constraint is satisfied, we get rid of this case by using Lemma 4 of [7], stating the following:

▶ **Lemma 1.** The problem of generating non-reachability certificates for an orbit instance $\mathcal{O}(A, X, Y)$ can be reduced to the problem of generating reachability certificates for an orbit instance $\mathcal{O}(A', X', Y')$ where $A'$ is invertible.

**Case 2: there exist real eigenvalues $\lambda$ and $|\lambda| \neq 1$.**

The key of the following property lies in [5], stating that $\lambda$-left eigenvectors $\varphi$ of a linear transformation $f$ are its invariants. More precisely, we can see that if $\varphi$ is a left-eigenvector of a linear transformation $A$, then by definition the following holds:

$$\forall v \in \mathbb{K}^d, \langle \varphi, Av \rangle = \lambda \langle \varphi, v \rangle \tag{1}$$

If $|\lambda| > 1$ (resp. $|\lambda| < 1$), then the sequence $(|\langle \varphi, A^n v \rangle|)$ (for $n \in \mathbb{N}$) is *strictly increasing* (resp. *strictly decreasing*),

▶ **Property 1.** Let $A \in \mathcal{M}_d(\mathbb{Q})$ a linear transformation and $\mathcal{O}(A, X, Y)$ an instance of the Orbit problem with no solution. Searching for a non-reachability certificate of an instance of the Orbit problem when $A$ admits real eigenvalues $\lambda$ such that $|\lambda| \neq 0$ and $|\lambda| \neq 1$ can be reduced to computing the eigenvector decomposition of $A$.

More precisely, if there exists $\varphi$ a $\lambda$-left-eigenvector of $A$ with $|\lambda| \neq 0$ and $|\lambda| \neq 1$, then there necessarily exists $N$ such that the couple $(N, P)$ defined as follows is a non-reachability certificate of $\mathcal{O}(A, X, Y)$.

1. If $|\langle \varphi, X \rangle| \neq 0$ and $|\langle \varphi, Y \rangle| = 0$, then $N = 0$ and $P = \{v : \langle \varphi, v \rangle \neq 0\}$
2. If $|\langle \varphi, X \rangle| = 0$ and $|\langle \varphi, Y \rangle| \neq 0$, then $N = 0$ and $P = \{v : \langle \varphi, v \rangle = 0\}$.
3. If $|\langle \varphi, X \rangle| \neq 0$ and $|\langle \varphi, Y \rangle| \neq 0$, $N = max(1, \lfloor \frac{ln(|\langle \varphi, Y \rangle|) - ln(|\langle \varphi, X \rangle|)}{ln(|\lambda|)} \rfloor + 1)$ and
   - If $|\lambda| > 1$, then $P = \{v : |\langle \varphi, v \rangle| \geqslant |\lambda. \langle \varphi, Y \rangle|\}$.
   - If $|\lambda| < 1$, then $P = \{v : |\langle \varphi, v \rangle| \leqslant |\lambda. \langle \varphi, Y \rangle|\}$.
4. Otherwise, if $d > 1$ there exist a transformation $B \in \mathcal{M}_{d-1}(\mathbb{Q})$ such that the problem of finding a certificate for $\mathcal{O}(A, X, Y)$ can be reduced to the problem of finding a certificate for $\mathcal{O}(B, X, Y)$.
   If $d = 1$, then $\mathcal{O}(A, X, Y)$ has a solution.

The certificate is semi-linear iff $\lambda \in \mathbb{Q}$.

**Proof.** Let $\varphi$ be a left-eigenvector of $A$ associated to the eigenvalue $\lambda$. We know that for all $v, \langle \varphi, v \rangle = k \Rightarrow \langle \varphi, Av \rangle = \lambda.k$. Let $U_n = |\langle \varphi, A^n X \rangle|$ be the $n$-th reachable state from $X$. If $|\lambda| < 1$ (resp. $|\lambda| > 1$), then $(U_n)$ is strictly decreasing (resp. strictly increasing).
1. Let $k_v = |\langle \varphi, v \rangle|$. If $k_X \neq 0$ and $k_Y = 0$, then the sequence $(U_n)$ never reaches $k_Y$, as for all $n$, $U_n \neq 0$. In other words, $|U_n| > 0$ for all $n \in \mathbb{N}$. Then it is clear that $P = \{X : |\langle \varphi, X \rangle| \neq 0\}$ is a valid certificate set of index $N = 0$.

2. Similarly, if $k_X = 0$ and $k_Y \neq 0$, then $P = \{X : |\langle \varphi, X \rangle| = 0\}$ and $N = 0$.
3. Assume now that $k_X \neq 0$ and $k_Y \neq 0$. If $k_X < k_Y$ and $|\lambda| < 1$ (respectively $k_X > k_Y$ and $|\lambda| > 1$), then $(1, \{v : |\langle \varphi, v \rangle| \leqslant |\lambda|.k_Y\})$ is a valid certificate set (respectively $(1, \{v : |\langle \varphi, v \rangle| \geqslant |\lambda|.k_Y\}))$. Otherwise, let us assume $|\lambda| < 1$ and $k_x \geq k_y$. $U_n$ is strictly decreasing, so there exist a $N$ such that $U_N \geqslant k_Y$ and $U_{N+1} < k_Y$. This implies that $Y$ can only be reachable after a finite number of iterations $N$. We also have that $U_{N+1} \geqslant |\lambda|.k_Y$ and $U_{N+2} < |\lambda|.k_Y$. If for all $n < N+1$, $Y \neq A^n X$, we can define $P = \{v : |\langle \varphi, v \rangle| < |\lambda|.k_Y\}$, and obtain $Y \notin P$ and $\{A^{N+1+n}X | n \in \mathbb{N}\} \subset P$. Therefore, the couple $(N+1, P)$ is a non-reachability certificate of $\mathcal{O}(A, X, Y)$. A similar proof for $|\lambda| > 1$ is valid as the sequence $U_n$ is now strictly increasing and the couple $(N, \{|\langle \varphi, X \rangle| \geqslant |\lambda|.k_Y\})$ is the corresponding certificate.

   We will now study the exact value of $N$. If $Y$ is reachable, then there exists a unique value of $N$ such that $|\lambda|^N |\langle \varphi, X \rangle| = k_Y$. This value is precisely $\frac{ln(|\langle \varphi, Y \rangle|) - ln(|\langle \varphi, X \rangle|)}{ln(|\lambda|)}$. If for every value of $n \leqslant N$, $Y$ is not reached and as $Y$ noes not belong to the certificate set $P$, the couple $(max(0, \lfloor N \rfloor), P)$ is a non-reachability certificate.
4. Assume $k_X = k_Y = 0$. In this case for every $n$, $\langle \varphi, A^n X \rangle = 0$, thus the linear combination of variables $\varphi.X$ is always equal to 0. There exists a base $\mathcal{B}$ of the transformation in which there exists a variable $v$ which remains null for every iteration of the transformation. In other words, there exist $A', Q$ such that $A' = Q.A.Q^{-1}$.

   Assume $d > 1$ and let $B' = A'_{|V \setminus v}$ and $Q' = Q_{|V \setminus v}$ the transformations restricted to all variables but $v$ (by removing both the associated line and column). Finding a certificate for $A$ is reduced to finding a certificate for $B = Q'^{-1}B'Q'$.

   If $d = 1$ and there exist a linear combination $\varphi$ of $X$ such that $\langle \varphi, X \rangle = 0$, then $X = 0$. Similarly, $Y = 0$.

Concerning the linearity of the certificate, if $\lambda \in \mathbb{Q}$, then every coefficient of $\varphi$ also belongs to $\mathbb{Q}$. Indeed $A$ has rational coefficients, so does $\varphi A = \lambda.\varphi$. Similarly, if $\varphi$ has rational coefficients, $\varphi.A = \lambda.\varphi$ also does.

In the case of $k_X \neq 0$ and $k_Y \neq 0$, we also have to get rid of the absolute value around $\langle \varphi, v \rangle$ in the definition of the certificate set. If $|\lambda| > 1$, the certificate set $\{v : (\langle \varphi, v \rangle \geqslant |\lambda \langle \varphi, Y \rangle|) \wedge (\langle \varphi, v \rangle \leqslant -|\langle \varphi, Y \rangle|)\}$ is semilinear. A similar set can be found for $|\lambda| < 1$. ◀

**Certificate index.** Being able to minimize the number of necessary unrollings to prove the non reachability is useful. In this regard, notice that the certificate index value $N$ of Theorem 1 is such that for every $n < N$, $\langle \varphi, A^n X \rangle \notin P$. In other words, it is minimal for its associated certificate set.

**Example.** Consider the Orbit Problem $\mathcal{O}(A, X, Y)$ with

$$A = \begin{pmatrix} 0 & 3 & 0 & 0 \\ -3 & 3 & 1 & 0 \\ 0 & 0 & 2 & 1 \\ 1 & 1 & 0 & 1 \end{pmatrix}$$

$A$ admits two real eigenvalues $\lambda_1 \approx 0.642$ and $\lambda_2 \approx 2.48$ respectively associated to the left-eigenvectors $\varphi_1 = (-0.522, 0.355, -0.261, 0.73)$ and $\varphi_2 = (0.231, -0.36, -0.749, -0.506)$. This is enough to build two preliminary certificate sets that only depend on $Y$ : $P_1 = \{v. | \langle \varphi_1, v \rangle | \leqslant \lambda_1. | \langle \varphi_1, Y \rangle |\}$ and $P_2 = \{v. | \langle \varphi_2, v \rangle | \geqslant \lambda_2. | \langle \varphi_2, Y \rangle |\}$. Those can be used for any initial valuation of $X$.

Let's now set $X = (1, 1, 1, 1)$ and $Y = (-9, -7, 28, 7)$. We have then

- $\langle \varphi_1, X \rangle = 0.302$ and $\langle \varphi_1, Y \rangle = 0.015$, so $N = 7$.
- $\langle \varphi_2, X \rangle = -1.384$ and $\langle \varphi_2, Y \rangle = -24.073$, so $N = 4$.

We can easily verify that for any $n \leqslant 7$, $A^n X \neq Y$, so the certificates $(7, P_1)$ and $(4, P_2)$ are sufficient to prove the non reachability of $Y$.

**Complex eigenvalues.** The treatment of complex eigenvalues can be reduced to the Case 1 by the *elevation* method described in [4]. The idea is simple : if variables evolves linearly (or affinely) then any monomial of those variables also evolves linearly (or affinely). For example, given $f(x) = x + 1$, then the new value of $x^2$ after application of $f$ is $(x+1)^2 = x^2 + 2x + 1$, which is an affine combination of $x^2$, $x$ and 1. $f$ can be *elevated* to the degree 2 by expressing this new monomial : $f_2(x_2, x) = (x_2 + 2x + 1, x + 1)$.

▶ **Definition 4.** Let $A \in \mathcal{M}_d(\mathbb{K})$. We denote $\Psi_k(A)$ the *elevation* matrix such that $\forall v = (v_1, ..., v_d) \in \mathbb{K}^n, \Psi_k(A).p(X) = p(A.X)$, with $p \in (\mathbb{K}[X]^k)$ a polynomial associating $X$ to all possible monomials of degree $k$ or lower.

By extension, we denote $\Psi_k(v)$ a vector $v$ elevated to the degree $k$.

**Remark.** This transformation has the advantage to linearize some polynomial mappings [4]. For example, the mapping $f(x, y) = (x + y^2, y + 1)$ is equivalent to $g(x, y, y_2, \mathbb{1}) = (x + y_2, y + \mathbb{1}, y_2 + 2y + \mathbb{1}, \mathbb{1})$ on multiple iterations given the right initial value of $y_2 = y^2$ and $\mathbb{1} = 1$. The whole class of solvable polynomial mappings [13, 4] is actually linearizable. We also have the following property [4]:

▶ **Property 2.** Let $A \in \mathcal{M}_d(\mathbb{Q}), \Lambda(M)$ the eigenvalue set of a matrix $M$ and $k$ an integer. Then for any product $p$ of $k$ or less elements of $\Lambda(A), p \in \Lambda(\Psi_k(A))$ where $\Psi_k(A)$ is the elevation of $A$ to the degree $k$.

The product of all eigenvalues is the determinant of the transformation, which is by construction a rational. The elevation to the degree $n$ where $n$ is the size of the matrix admits then at least one rational eigenvalue. We can deduce from this the following theorem.

▶ **Theorem 1.** Let $\mathcal{O}(A, X, Y)$ be an unsatisfiable instance of the Orbit problem with $A \in \mathcal{M}_n(\mathcal{Q})$ admitting at least one eigenvalue $\lambda \in \mathbb{C}$ such that $|\lambda| \neq 0$ and $|\lambda| \neq 1$. Then left eigenvectors of $\Psi_d(A)$ provide :
- real linear semialgebraic certificates for $d = 1$ ($\Psi_1(A) = A$) if there exist real eigenvalues;
- real semialgebraic certificates of degree 2 for $d = 2$ if there exist complex eigenvalues;
- at least one rational certificate of degree $n$ for $d = n$ if $|det(A)| \neq 1$.

**Proof.** We treat each case separately:
- The case where $A$ admits real eigenvalues is treated by Property 1;
- If $A$ admits a complex eigenvalue $\lambda$, $A$ also admits its conjugate $\bar{\lambda}$ as eigenvalue. By Property 2, $\Psi_2(A)$ admits $\lambda.\bar{\lambda}$ as a real eigenvalue, which is treated by Property 1;
- The product of all eigenvalues of a rational matrix is rational. As such, $\Psi_n$ necessarily admit a rational eigenvalue which implies the existence of an associated rational eigenvector that can be used, according to Property 1, as a certificate.

◀

**Remark.** The image of $A \in \mathcal{M}_d(\mathbb{K})$ is a projection of the image of $\Psi_k(A)$ for any $k$, and semialgebraic certificates of $A$ are, by extension, semilinear certificates of $\Psi_n(A)$. The size of $\Psi_k(A)$ is $\binom{d+k}{k}$, which is $O(d^2)$ when $k = 2$ and $O(d^d)$ when $d = k$. An eigenvector computation has a polynomial time complexity (slightly better than $O(d^3)$). The two first cases of Theorem 1 are thus computable in polynomial time in the number of variables.

**Example.** The matrix from the previous example admits two complex eigenvalue $\lambda \approx 1.439 + 2.712i$ and $\bar{\lambda}$. As $\lambda\bar{\lambda} \approx 9.425$, it also admits a polynomial invariant $\varphi$ (whose size is too long to fit in this article as it manipulates 10 monomials). However, $\langle \varphi, X \rangle = 0.220$ and $\langle \varphi, Y \rangle = 195.738$, thus the associated index is 4.

**Case 3: all eigenvalues have a modulus equal to** $1$

This case is trickier as eigenvectors do not give information about the convergence or the divergence of the linear combination of variables they represent. For example, let us study the orbit problem $\mathcal{O}(A, X, Y)$ where $A$ is the matrix associated with the mapping $f(x, \mathbb{1}) = (x + 2*\mathbb{1}, \mathbb{1})$, $X = (0, 1)$ and $Y = (5, 1)$. $x_Y$ is odd, thus $Y$ is not reachable. f admits only $\varphi = (0, 1)$ as left-eigenvector associated to the eigenvalue $\lambda = 1$, meaning that $\langle (0, 1), (x, \mathbb{1}) \rangle = \langle (0, 1), f(x, \mathbb{1}) \rangle$ for any $x$. As $\langle (0, 1), (x, \mathbb{1}) \rangle = \mathbb{1}$, we are left with the invariant $\mathbb{1} = 1$. This invariant is clearly insufficient to prove that $Y$ is not reachable.

$f$ thankfully admits a generalized left-eigenvector $\mu = (\frac{1}{2}, 1)$ associated to 1. More precisely, $\mu A = \mu + \varphi$, which implies that $\mu A^n X = (\mu + n\varphi).X$. In other words, we have $\frac{1}{2}x + 1 = \frac{1}{2}x_X + 1 + n$ which simplifies into $\frac{1}{2}x = n$. The couple $(3, \{(x, y) : \exists n > 3, \frac{1}{2}x = n\})$ is a non reachability certificate.

▶ **Property 3.** Let $A$ a linear transformation and $\{e_i\}_{i<N}$ $N$ linked 1-left eigenvectors (i.e. $e_0 A = e_0$ and for $0 < i < N$, $e_i A = e_i + e_{i-1}$). Then for all $i < N$, $\langle e_i A^k, X \rangle = P_i(k, X)$, where $P_i(k, X)$ is a polynomial of degree $i$ in the variable $k$ and 1 in each variable of $X$.

**Proof.** Let $\{e_i\}_{i<N}$ a family of $N$ linked 1-left eigenvectors. We can compute $P_i(k, X)$ by induction on $i$. For $i = 0$, $e_0$ verifies $e_0 A^k = e_0 = \binom{k}{i} e_{N-i-1}$. Assume now $e_i A^k = P_i(k)$ are vectors of polynomials of degree at most $i$. Then, we have $e_{i+1}.A^{k+1} = (e_{i+1} + e_i).A^k = e_{i+1}A^k + P_i(k)$ Now, let $U_{k+1} = U_k + P_i(k)$. Then $U_k = U_0 + \sum_{l=0}^{k} P_i(l)$ is a vector of polynomials of degree at most $i + 1$. ◀

Thus, there exists a linear combination of variables of $X$ that diverges. This is enough to certify the non reachability of the Orbit Problem *for non diagonalizable matrices with the eigenvalue $\lambda = 1$.*

**Remark.** Even if the first eigenvector is enough to represent a non-reachability certificate, every generalized eigenvector also can. By property 3, the value of the linear combination described by a generalized eigenvector $\varphi$ evolves polynomially, thus it eventually always decrease or increase (after the highest root of its derivate). That is why for a given objective $Y$ there exist a finite number of $n$ such that $|\varphi Y| \leqslant |\varphi A^n X|$, thus after this $n$, $\{v : |\varphi v| > |\varphi Y|\}$ is a certificate.

**Complex eigenvalues.** If $\lambda \in \mathbb{C}$, we will use the same trick we used for complex eigenvalues of Case 2. As for every complex eigenvalue $\lambda$ of $A$, $\bar{\lambda}$ is also an eigenvalue, then $\lambda.\bar{\lambda} = 1$ is an eigenvalue of $\Psi_2(A)$ by property 2. Thus :

▶ **Theorem 2.** Let $\mathcal{O}(A, X, Y)$ be a non satisfiable instance of the Orbit Problem such that for all eigenvalue $\lambda$ of $A$, $|\lambda| = 1$ and $A$ is not diagonalisable. Then there exist a family of 1-left-eigenvectors $\mathcal{F} = \{e_0, ..., e_n\}$ of $\Psi_2(A)$ such that for all $1 \leqslant i \leqslant n$, $Q_i(n) = \langle e_i, \Psi_2(A)^n \Psi_2(X) \rangle$ is a polynomial and $(N, P)$ is a non reachability certificate with:

- $N = \lfloor max(\{0\} \cup \{x \in \mathbb{R}.Q(x) = \langle e_i, \Psi_2(A^x)\Psi_2(Y) \rangle \}) \rfloor$

- $P = \{v : |\langle e_i, \Psi_2(A)^n \Psi_2(v) \rangle | \geqslant |Q(N)|\}$

**Proof.** Let $\mathcal{O}(A, X, Y)$ be an instance of the Orbit Problem. We will reduce the problem to the case where $A$ has positive rational eigenvalues, i.e. $\lambda = 1$ and $A$ admits a family $\mathcal{F}$ of left-eigenvectors of size $|\mathcal{F}| > 1$. In this case, by Property 3 we know that there exists a linear combination of variables $v$ following a polynomial evolution described by $Q$ such that $deg(Q) > 0$. As $Q$ eventually diverges, there exists a $N$ such that for all $N' > N$, $|v(A^{N'}X)| > |v(Y)|$. This $N$ is the maximum between 0 and the highest value of $x$ such that $Q(x) = v(Y)$ as, for any higher value of $x$, $|Q(x)| > |v(Y)|$. Also, the set $\{v.|\langle e_i, \Psi_2(A)^n \Psi_2(v) \rangle | \geqslant |Q(N)|\}$ contains all reachable configurations but does not contain $Y$, thus $(N, P)$ is a valid certificate.

In the general case where $\lambda \in \mathbb{C}$, we will use Property 2 to show that if there exist complex eigenvalues $\lambda$ such that $|\lambda| = 1$, of multiplicity $m > 1$ with $m \neq dim(ker(A - \lambda Id))$, then $\Psi_2(A)$ admits 1 or $-1$ as an eigenvalue and its multiplicity $m' > 1 \neq dim(ker(\Psi_2(A) - \lambda Id))$. This implies directly the existence of generalized eigenvector, thus of a family of left-eigenvectors of size strictly higher than 1. To this purpose, we refer to basic properties of $\Psi_d$:

▶ **Lemma 2.**
1. $\Psi_k(A.B) = \Psi_k(A).\Psi_k(B)$
2. $\Psi_k(A^{-1}) = \Psi_k(A)^{-1}$

**Proof.** 1. $\Psi_k(A).\Psi_k(B)p(X) = \Psi_k(A).p(B.X) = p(A.B.X) = \Psi_k(A.B)p(X)$
2. $\Psi_k(A^{-1}).\Psi_k(A).p(X) = p(A.A^{-1}X) = p(X)$ so $\Psi_k(A^{-1}).\Psi_k(A) = Id$.

◀

Let $J$ the Jordan normal form of $A$, i.e. there exists $P$ such that $A = P^{-1}JP$. We have that $J = \begin{pmatrix} J_1 & 0 & \ldots & 0 \\ 0 & \ddots & \ddots & \vdots \\ \vdots & \ddots & \ddots & 0 \\ 0 & \ldots & 0 & J_k \end{pmatrix}$, and $J_k = \begin{pmatrix} \lambda_k & 1 & \ldots & 0 \\ 0 & \ddots & \ddots & \vdots \\ \vdots & \ddots & \ddots & 1 \\ 0 & \ldots & 0 & \lambda_k \end{pmatrix}$

From Lemma 2, it is easy to prove that $\Psi_d(A) = \Psi_d(P)^{-1}\Psi_d(A)\Psi_d(P)$. As $\Psi_d(A)$ and $\Psi_d(J)$ are similar, they have the same eigenvalues. We know that there exist $v_1, v_2, v_3$ in the base of $J$ such that

- $v'_1 = \lambda.v_1 + v_2$
- $v'_2 = \lambda.v_2$
- $v'_3 = \bar{\lambda}.v_3$

where $v'_i$ is the new value of $v_i$ in the base of $J$. Then the image of $v_1v_3$ (denoted $(v_1v_3)'$) with respect to $\Psi_2(J)$ is $v_1v_3 + \bar{\lambda}.v_2.v_3$. Also, we know that $(v_2v_3)' = v_2v_3$. Let $\varphi$ such that $\varphi.\Psi_2(J).V = v_1v_3$.

$$\begin{aligned} \varphi.(\Psi_2(J) - Id)V &= v_1v_3\bar{\lambda}.v_2v_3 - v_1v_3 \\ &= \bar{\lambda}.v_2v_3 \\ \varphi.(\Psi_2(J) - Id)^2V &= \bar{\lambda}.v_2v_3 - \bar{\lambda}.v_2v_3 \\ \varphi.(\Psi_2(J) - Id)^2V &= 0 \end{aligned}$$

As this is true for any $V$, then $\varphi.(\Psi_2(J) - Id) \neq 0$ and $\varphi.(\Psi_2(J) - Id)^2 = 0$. In conclusion, $\varphi$ is a generalized eigenvector of $\Psi_2(J)$, thus $\Psi_2(A)$ also admits a generalized eigenvector.

◀

**Example.** We consider the Orbit problem $\mathcal{O}(A, X, Y)$ with $A = \begin{pmatrix} 1 & 1 & 0 \\ 0 & 1 & 1 \\ 0 & 0 & 1 \end{pmatrix}$, $X = (-2, -1, 1)^t$ and $Y = (2, 6, 1)^t$. $A$ admits as 1-generalized-left-eigenvectors: $\{e_0 = (0, 0, 1); e_1 = (0, 1, 0); e_2 = (1, 0, 0)\}$. By the previous property, we know that $e_2 A^k = e_2 + k.e_1 + \frac{k(k-1)}{2}.e_0$, thus

$$\begin{aligned} \langle e_2 A^k, (x_X, y_X, \mathbb{1}) \rangle &= y_X + k x_X + \frac{k(k-1)}{2} \\ &= \frac{1}{2} k^2 - \frac{5}{2} k - 1 \end{aligned}$$

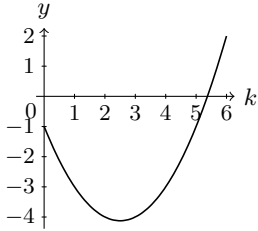

**Figure 1** Graph of the polynomial y $= \frac{1}{2} k^2 - \frac{5}{2} k - 1$

As we can see in Figure 1, from $k = 3$, the value of $x$ is strictly increasing and after $k = 7$, the value of $x$ is strictly superior to 2. Thus we have to check a finite number of iterations before reaching $x > 2$, which is the certificate set constraint of the non-reachability of $Y$. For $k \in [0, 6]$, $Y$ is not reached. The couple $(7, \{(x, y, \mathbb{1}).x > 2\})$ is thus a certificate of non reachability of $Y$.

**Case 4: eigenvalues all have a modulus equal to $1$ and the transformation is diagonalizable**

Some transformations do not admit generalized eigenvectors, namely diagonalizable transformations. The previous theorem is then irrelevant if for every eigenvalue $\lambda$, $|\lambda| = 1$. Such transformations are *rotations* : they remain in the same set around the origin. Take as example the transformation $A$ of Figure 2, taken from [7].

It defines a counterclockwise rotation around the origin by angle $\theta = \arctan(\frac{3}{5})$, and $\frac{\theta}{\pi}$ is not rational. The reachable set of states from $X$, i.e. $\{X, AX, A^2 X, ...\}$ is strictly included in its closure, i.e. the set of reachable states and their neighbourhood. As $Y$ is not on the closure of the set, then we can easily provide a non-reachability semi-algebraic invariant certificate of $Y$, that is the equation of the circle. However, we cannot give such a certificate for $Z$ though it is not reachable. If it were reachable, there would exist a $n$ such that $A^n X = Z$, thus $A^{2n} X = X$. $n$ would also satisfy $\theta * n = 0[2\pi]$, which is impossible as $\frac{\theta}{\pi}$ is not rational. More generally, the closure of the reachable set of states of diagonalisable transformations with eigenvalues of modulus 1 is a semialgebraic set [7]. Semialgebraic certificates for such transformations exist if and only if $Y$ does not belong to this closure [7].

▶ **Theorem 3.** *For a given instance $\mathcal{O}(A, X, Y)$ such that $A$ is diagonalizable and all its eigenvalues have a modulus of 1, eigenvectors can be used as semialgebraic certificates iff $Y$ is not in the closure.*

$$\begin{aligned} A &= \tfrac{1}{5} \begin{pmatrix} 4 & -3 \\ 3 & 4 \end{pmatrix} \\ X &= (1, 0) \\ Y &= (1.5, 0.7) \\ Z &= (-1, 0) \end{aligned}$$

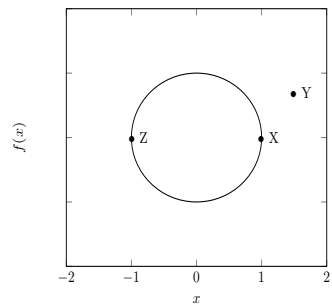

**Figure 2** Closure of the reachable set of $A$ starting with $X$.

**Proof.** Let $\mathcal{O}(A, X, Y)$ be an instance of the Orbit Problem with $A$ a diagonalizable matrix only admitting eigenvalues $\lambda$ such that $|\lambda| = 1$ Let $\varphi$ an eigenvector of $A$, we denote $R = \{v | \exists k. A^k X = v\}$ the reachable set.

▶ **Lemma 3.** Let $(\lambda_i, \varphi_i)$ be $d$ couples of eigenvalue / left-eigenvector of a diagonalizable matrix $A$ of size $d$. Then $R = \{v | \exists k, \forall 1 \geqslant i \geqslant d, \varphi_i.v = \lambda_i^k X\}$

**Proof.** Let $R' = \{v | \exists k, \forall 1 \geqslant i \geqslant d, \varphi_i.v = \lambda_i^k X\}$. By the definitions of $R$ and $\varphi_i$, the inclusion $R \subset R'$ is trivially true. Now take $v \in R'$. As there exist $d$ different and independent eigenvectors, $v$ is a solution of the following relation: $\exists k. \Phi v = (\lambda_1^k x_1, ... \lambda_d^k x_d)^t$, where $\Phi$ is an invertible matrix whose lines are directly defined by eigenvectors. As $\Phi$ is invertible, there exists only one solution for each $k$. As $v$ is one of those solutions, then $v \in R$. ◀

By lemma 3, for any $i$ between 1 and $d$, every element $v$ of $R$ verifies $|\langle \varphi_i, v \rangle| = |\langle \varphi_i, X \rangle|$, thus $R \subset R_\varphi = \{v : |\langle \varphi_i, v \rangle| = |\langle \varphi_i, X \rangle|\}$. Note that this inclusion is strict, as $X' = A^{-1}X \in R_\varphi$ but $X' \notin R$. If $Y$ does not belong to $R_\varphi$, then $(0, R_\varphi)$ is a non reachability certificate. ◀

## 3.2 General existence of a certificate for the integer Orbit Problem

The Orbit Problem is originally defined on $\mathbb{Q}$, but most programs only work on integers. Though $\mathbb{Z}$ is not a field, it is still possible to define linear transformations on $\mathbb{Z}$. Basic matrix operations involving divisions (such as inversion) are forbidden, but the only relevant operation in our case is multiplication (does there exist a $n$ such that $A^n X = Y$ ?) which is consistent for integer matrices.

When dealing with linear transformations manipulating integers, things are quite different. Indeed, the following property holds for integer matrices.

▶ **Property 4.** Let $A \in \mathcal{M}_n(\mathbb{Z})$. If all its eigenvalue $\lambda$ have a modulus inferior or equal to 1, then there exists $n > 1$ such that $\lambda^n = \lambda$.

**Proof.** Let $A \in \mathcal{M}(\mathbb{Z})$ such that for all eigenvalue $\lambda$, $|\lambda| \leqslant 1$.

If $\lambda = 0$, then we can conclude right away ($0^2 = 0$).

The characteristic polynomial $P \in \mathbb{Z}[X]$ of $A$ is monic, i.e. its leading coefficient is 1. Thus by definition, every eigenvalue is an algebraic integer. We will use the Kronecker theorem [14], stating that if a non null algebraic integer $\alpha$ has all its rational conjugates (i.e. roots of its rational minimal polynomial) admitting a modulus inferior or equal to 1, then $\alpha$ is a root of unity.

Each eigenvalue $\lambda$ admits a minimal rational polynomial $Q$. We can show that $Q$ necessarily divides $P$ by performing an euclidian division : there exist $D, R \in \mathbb{Q}[X]$ such that $P(X) = Q(X)D(X) + R(X)$, with the degree of $R$ strictly inferior to $Q$. We know that $P(\lambda) = 0$ and $Q(\lambda) = 0$, thus $R(\lambda) = 0$. If $R \neq 0$, then $R$ is the minimal polynomial of $\lambda$ as its degree is inferior to the degree of $Q$, which is absurd by hypothesis. Thus, the set of rational conjugates of $\lambda$ are roots of $P$, by hypothesis of modulus inferior or equal to 1. By the Kronecker theorem, $\lambda$ is a root of unity, i.e. $\exists n > 1. \lambda^n = \lambda$.

◀

This result is fundamental in the proof of the following theorem.

▶ **Theorem 4.** Any non-reachable instance of the Orbit problem $\mathcal{O}(A, X, Y)$ where $A \in \mathcal{M}_n(\mathbb{Z})$ admit a closed semi-algebraic invariant.

**Proof.** We already treated the case where the matrix has an eigenvalue whose modulus is different from 1 (Property 1) and the case where the matrix is not diagonalizable (Property 3). We are left with the hypothesis of the Property 4.

Let $A$ be a transformation such that all its eigenvalue are either 0 or roots of unity. $A$ represents a finite-monoïd transformation, i.e. its reachable set of space is *finite*. More precisely, there exist $N, p$ such that $\forall n > N, A^{n+p} = A^n$ Let $P = \{A^N X, A^{N+1} X, ..., A^{N+p-1} X\}$. If $Y$ is not reachable, then the couple $(P, N)$ is a non-reachability certificate.

The closure of such a certificate comes from the same eigenvalue argument. The only case we had a non-closed certificate comes from Property 1 when $|\lambda| \neq 0$, $|\lambda| \neq 1$, $|\langle \varphi, X \rangle| \neq 0$ and $|\langle \varphi, Y \rangle| \neq 1$. As we also have $|\lambda| \geqslant 1$ for integer matrices, the certificate set $\{v : |\langle \varphi, v \rangle| \geqslant |\langle \varphi, X \rangle|\}$ is a valid closed certificate set. ◂

## 4 Conclusion and future work

This paper presents new insights on the quality of certificates necessary to prove the non-reachability of a given Orbit problem instance. In addition, in contrast with [7], we gain simplicity and precision by not studying the Jordan normal form of a linear transformation but only its eigenvector decomposition.

Certificate sets of transformations of the two first cases treated in Section 3.1($|\lambda| \neq 1$) are totally independent of the initial state $X$, which widens the possible uses of certificates. It is possible to use the same certificate set for differents values of $X$ and $Y$, allowing to treat specific kind of vector sets (coefficients of $X$ and $Y$ as closed intervals for example, which are encountered more often in program verification than precise values). Interesting axis of development are to find certificates independent of $X$ and $Y$ in the general case and to study in detail which kind of vector sets can the certificate search be of use.

As this article explores the Orbit Problem for rationals, it is worth noting that certificates may not necessarily be relevant for real-life programs manipulating floats. For example, the Orbit problem $(x \mapsto \frac{x}{2}, 1, 0)$ has a solution for some floating point implementations due to limited precision. The question of certificates synthesis for such problems is also an interesting challenge.

## References

1   Sandrine Blazy, David Bühler, and Boris Yakobowski. Structuring abstract interpreters through state and value abstractions. In *VMCAI 2017, Proceedings*, pages 112–130, 2017.
2   Marius Bozga, Radu Iosif, and Filip Konecný. Fast acceleration of ultimately periodic relations. In *Computer Aided Verification, 22nd International Conference, CAV 2010, Edinburgh, UK, July 15-19, 2010. Proceedings*, volume 6174 of *Lecture Notes in Computer Science*, pages 227–242. Springer, 2010.
3   Patrick Cousot and Radhia Cousot. Abstract interpretation: a unified lattice model for static analysis of programs by construction or approximation of fixpoints. In *Proceedings of the 4th ACM SIGACT-SIGPLAN symposium on Principles of programming languages*, pages 238–252. ACM, 1977.
4   Steven de Oliveira, Saddek Bensalem, and Virgile Prevosto. Polynomial invariants by linear algebra. In *International Symposium on Automated Technology for Verification and Analysis*, pages 479–494. Springer, 2016.
5   Steven de Oliveira, Saddek Bensalem, and Virgile Prevosto. Synthesizing invariants by solving solvable loops. In *Automated Technology for Verification and Analysis - 15th International Symposium, ATVA 2017, Pune, India, October 3-6, 2017, Proceedings*, pages 327–343, 2017.

**6**   Michael D Ernst, Jake Cockrell, William G Griswold, and David Notkin. Dynamically discovering likely program invariants to support program evolution. *IEEE Transactions on Software Engineering*, 27(2):99–123, 2001.
**7**   Nathanaël Fijalkow, Pierre Ohlmann, Joël Ouaknine, Amaury Pouly, and James Worrell. Semialgebraic Invariant Synthesis for the Kannan-Lipton Orbit Problem. In *34th Symposium on Theoretical Aspects of Computer Science, STACS 2017, March 8-11, 2017, Hannover, Germany*, volume 66 of *LIPIcs*, pages 29:1–29:13. Schloss Dagstuhl - Leibniz-Zentrum fuer Informatik, 2017.
**8**   Jean-Christophe Filliâtre and Andrei Paskevich. Why3—where programs meet provers. In *European Symposium on Programming*, pages 125–128. Springer, 2013.
**9**   Ravindran Kannan and Richard J Lipton. The orbit problem is decidable. In *Proceedings of the twelfth annual ACM symposium on Theory of computing*, pages 252–261. ACM, 1980.
**10**  Ravindran Kannan and Richard J Lipton. Polynomial-time algorithm for the orbit problem. *Journal of the ACM (JACM)*, 33(4):808–821, 1986.
**11**  Laura Kovács. Reasoning algebraically about P-solvable loops. In *International Conference on Tools and Algorithms for the Construction and Analysis of Systems*, pages 249–264. Springer, 2008.
**12**  Herbert Rocha, Hussama Ismail, Lucas Cordeiro, and Raimundo Barreto. Model checking embedded c software using k-induction and invariants. In *Computing Systems Engineering (SBESC), 2015 Brazilian Symposium on*, pages 90–95. IEEE, 2015.
**13**  Enric Rodríguez-Carbonell and Deepak Kapur. Generating all polynomial invariants in simple loops. *J. Symb. Comput.*, 42(4):443–476, 2007.
**14**  Andrzej Schinzel, Hans Zassenhaus, et al. A refinement of two theorems of kronecker. *Michigan Math. J*, 12:81–85, 1965.